\documentstyle[epsfig,12pt]{JHEP}
\preprint{LPT-Orsay 00-30}

\title{Perturbative and non-perturbative aspects of moments of the
thrust distribution in $e^+e^-$ annihilation\footnote{Research 
supported by the EC program ``Training and Mobility of Researchers'', Network 
``QCD and Particle Structure'', contract ERBFMRXCT980194.}}

\author{
{\bf Einan Gardi}\\ \vspace{.3in} 
Laboratoire de Physique Th\'eorique\footnote{CNRS UMR 8627},
Universit\'e de Paris XI, \\
91405 Orsay Cedex, France
\\{\rm e-mail:} { \tt Einan.Gardi@th.u-psud.fr}
}

\abstract{Resummation and power-corrections play a crucial role in the
phenomenology of event-shape variables like the thrust~$T$. 
Previous investigations showed that the perturbative contribution to 
the average thrust is dominated by gluons of small invariant mass, of 
the order of~$10\%$ of~$Q$, where~$Q$ is the center-of-mass energy. The effect
of soft gluons is also important, leading to a 
non-perturbative~$1/Q$~correction. 
These conclusions are based on renormalon analysis in the 
single dressed gluon (SDG) approximation. 
Here we analyze higher moments of the thrust distribution using a
similar technique. We find that the 
characteristic gluon invariant mass contributing 
to~${\large<}(1-T)^m{\large>}$ increases with~$m$. 
Yet, for~$m=2$ this scale is quite low, around $27\%$
of~$Q$, and therefore renormalon resummation is still very
important. On the other hand, the power-correction 
to~${\large<}(1-T)^2{\large>}$ from a single soft gluon emission is 
found to be highly suppressed: it scales as~$1/Q^3$. In practice, 
${\large<}(1-T)^2{\large>}$ and higher moments depend also on soft gluon 
emission from configurations of three hard partons, which may 
lead to $\alpha_s(Q^2)/Q$ power-corrections. This issue is yet to 
be investigated.
}


\newcommand{\mycomm}[1]{\hfill\break
 $\phantom{a}$\kern-3.5em{\tt===$>$ \bf #1}\hfill\break}
\newcommand{\mycommA}[1]{\hfill\break
$\phantom{a}$\kern-3.5em{\tt***$>$ \bf #1}\hfill\break}

\begin{document}
\catcode`\@=11 
\def\lsim{\mathrel{\mathpalette\@versim<}}
\def\gsim{\mathrel{\mathpalette\@versim>}}
\def\@versim#1#2{\vcenter{\offinterlineskip
        \ialign{$\m@th#1\hfil##\hfil$\crcr#2\crcr\sim\crcr } }}
\catcode`\@=12 
\def\beq{\begin{equation}}
\def\eeq{\end{equation}}
\def\MSbar {\hbox{$\overline{\hbox{\tiny MS}}\,$}}
\def\eff{\hbox{\tiny eff}}
\def\FP{\hbox{\tiny FP}}
\def\PV{\hbox{\tiny PV}}
\def\SF{\hbox{\tiny SF}}
\def\IR{\hbox{\tiny IR}}
\def\UV{\hbox{\tiny UV}}
\def\ECH{\hbox{\tiny ECH}}
\def\APT{\hbox{\tiny APT}}
\def\QCD{\hbox{\tiny QCD}}
\def\CMW{\hbox{\tiny CMW}}
\def\pinch{\hbox{\tiny pinch}}
\def\brem{\hbox{\tiny brem}}
\def\V{\hbox{\tiny V}}
\def\BLM{\hbox{\tiny BLM}}
\def\NLO{\hbox{\tiny NLO}}
\def\SDG{\hbox{\tiny SDG}}
\def\soft{\hbox{\tiny large-angle}}
\def\res{\hbox{\tiny res}}
\def\PT{\hbox{\tiny PT}}
\def\PA{\hbox{\tiny PA}}
\def\1loop{\hbox{\tiny 1-loop}}
\def\2loop{\hbox{\tiny 2-loop}}
\def\mysim{\kern -.1667em\lower0.8ex\hbox{$\tilde{\phantom{a}}$}}
\def\a{\bar{a}}

\section{Introduction}

The systematic analysis  of event-shape variables 
in $e^+e^-$ annihilation has become an active research field in
the recent years~\cite{Dokshitzer:1999ai}. 
High quality experimental data in a wide range of center-of-mass 
energies $Q$ (from $10$ up to $200\,{\rm GeV}$) provide an 
opportunity for precision test of the theory. On the theoretical 
side, the simple set-up of $e^+e^-$ annihilation 
allows to probe directly the QCD vacuum and learn about the interface
between perturbative and non-perturbative physics.

Empirically, it is known for quite some time \cite{Pluto} that perturbative
approximations to moments of event-shape variables are consistent with
experimental data only upon including additive power-corrections.
A classical example is the case of the thrust, defined by
\beq
T=\frac{\sum_i \left\vert \vec{p}_i \cdot \vec{n}_T\right\vert}
{\sum _i \vert \vec{p}_i \vert }
\label{T_def}
\eeq
where the summation is over all the final-state particles $\vec{p}_i$. 
The thrust axis $\vec{n}_T$ is a unit vector 
which is set, for a given event, such that $T$ is maximized. 
Below we shall use the variable $t\equiv 1-T$ which vanishes in the 
two-jet limit. 

For the average thrust, as for several other average event-shape
variables, one finds $1/Q$ corrections,
${\large<}t{\large>}\simeq{\large<}t{\large>}_{\PT}+\lambda/Q$, where
$\lambda$ is some hadronic scale. These corrections are quite large at LEP
energies.
The case of average event-shape variables was intensively studied both
theoretically~\cite{W}--\cite{Shape_function2} and experimentally.

Here we investigate perturbative and non-perturbative aspects of 
higher moments of $1\,-$ thrust, ${\large<}t^m{\large>}$. 
This subject was addressed before
(see e.g. \cite{Shape_function3,Shape_function4,Webber:1999zj,Beneke}),
but it did not receive yet the appropriate attention.
Whereas some experimental results are already 
available~\cite{L3,Wicke:1998fq}, 
many more can be extracted from stored LEP data \cite{Biebel}. 
This analysis is worthwhile: alongside interesting
theoretical insights, the investigation of higher moments of event-shape 
variables may lead to additional precision measurements of $\alpha_s$.

Existence of a well-defined perturbative approximation to 
event-shape variables is guaranteed since they are infrared and 
collinear safe. However, the remaining sensitivity of these observables 
to soft and collinear emission is usually quite high and it shows up 
in the form of large perturbative coefficients.
For event-shape distributions close to the two-jet limit (the
Sudakov region) the leading logarithms in the perturbative coefficients 
can be resummed to all-orders~\cite{CTTW}. 
Out of this special part of phase-space, and in particular when moments 
of event-shape variables are considered, only the leading and
next-to-leading perturbative terms are known~\cite{Ellis,EVENT,EVENT2}.    
Since the apparent convergence of these perturbative 
expansions in a standard renormalization-scheme and scale is slow,
resummation seems necessary. 
A relevant source of large coefficients are renormalon diagrams 
\cite{DMW,thrust,AZ,Beneke,BBM,Shape_function1}
which reflect the effect of the running-coupling.
 
One can imagine~\cite{thrust,conf}, in analogy with the skeleton 
expansion in the
Abelian theory, a possibility of reorganizing the perturbative expansion 
such that all diagrams which correspond to dressing a single gluon are
summed first,
then diagrams with two dressed gluons, etc. Formally, a systematic
expansion of this type has not yet been shown to exist. However, the spirit
of the Brodsky-Lepage-Mackenzie (BLM) scale setting method \cite{BLM},
one can attempt to identify and resum the single dressed gluon~(SDG) 
terms to all orders. This is the basis of the renormalon resummation 
methodology, which was applied in many
different QCD examples~\cite{Beneke}. In the context of event-shape
variables in $e^+e^-$ annihilation we mention the
case of the longitudinal cross-section~\cite{BBM} and that of the average 
thrust~\cite{thrust}. 

Whatever resummation procedure is applied, it is clear 
that perturbation theory alone
cannot predict the observed values of event-shape variables: the
perturbative calculation uses quark and gluon fields while
the measurement is of hadrons.
The effect of hadronization on the observables can be modeled by
Monte-Carlo simulations. However, in order to gain some 
understanding on the nature of confinement in QCD it is favorable to
analyze raw hadronic data, and parameterize non-perturbative effects in
the simplest possible way.  

It was noticed that hadronization does not involve 
large momentum transfer. Therefore perturbative results calculated in
terms of partons can almost be directly compared with the 
data~\cite{Dokshitzer:1999ai}. 
Due to the sensitivity of the
observables to soft physics some modification of the perturbative result
is still necessary. At the perturbative level, infrared sensitivity
shows up in the form of infrared renormalons~\cite{Beneke}: 
the coefficients increase
fast, asymptotically as $n!$, and have a constant sign
pattern. Consequently, the series is non-summable. Different resummation
prescriptions, or ``regularizations'' of the perturbative sum, differ by
power-suppressed terms. This ambiguity must be compensated at the
non-perturbative level.
In the SDG approximation, one can calculate the perturbative sum and
extract the form of its ambiguity. This way~\cite{DMW,AZ,Shape_function1}
a perturbative calculation can be used to identify the parametric 
dependence of non-perturbative corrections on the external scale $Q$.
The magnitude of these power-corrections cannot be calculated, but assuming 
some universality properties~\cite{DMW,AZ} it is possible to estimate it 
for a large class of observables based on experimental data for one of them.

The SDG approximation, as applied in \cite{thrust} in the case of the
average thrust, provides a systematic framework for the
analysis of perturbative running-coupling effects together with the related 
power-corrections. These two aspects of improving the truncated
perturbative result naturally complement each other: 
they reflect the same physical phenomenon.

It is clear that any observable, including the moments of $1\,-$ thrust
considered here, may eventually depend on other non-perturbative
effects, which are unrelated to ambiguities of perturbation theory. 
We assume that these effects are not important.
Since the power-corrections identified from the ambiguities of the
perturbative expansion have a definite $Q$ dependence, this assumption 
can be confronted with experimental data.

It should be emphasized that, contrary to the Operator
Product Expansion (OPE), the SDG renormalon approach lacks the rigor of 
a systematic expansion: there is no small parameter which distinguishes 
the contribution of multiple emission from that of a single emission.
One should be aware of the possibility that in certain cases 
the leading power-corrections cannot be analyzed in the framework of the
SDG calculation as they are associated with soft gluon emission from 
configurations of three of more hard partons. 
In particular, in the case of the higher moments 
of $1\,-$ thrust, such effects may yield
$\alpha_s(Q^2)/Q$ 
power-corrections~\cite{Beneke,DW_dist,Shape_function3,Shape_function4}. 
We shall return to this important issue in the conclusions.

Further subtleties are related to the fact that event-shape variables
are not completely inclusive~\cite{Nason_Seymour,thrust}: 
they are sensitive to certain details of
the final state, while the resummation procedure we use is
completely inclusive with respect to the fragmentation products of the
gluon. 

The purpose of this work is to examine, in the framework of the SDG
approximation, running-coupling effects and power-corrections to 
higher moments of $1\,-$ thrust. 
We begin, in section 2 by analyzing the thrust distribution.
As in~\cite{thrust}, where the average thrust was analyzed, we use
the ``massive'' gluon dispersive approach \cite{DMW}. In this framework 
resummation as well as parameterization of power-corrections are obtained
from a so-called characteristic function. After explaining the main
assumptions of the SDG model we calculate the characteristic function 
for the thrust distribution. We then devote a short discussion 
to the domain of applicability of the SDG result as a function of the thrust. 
Next, in section 3 we evaluate the characteristic functions for various
moments of $1\,-$ thrust and study their properties. In particular we 
extract the characteristic invariant mass of gluons contributing to 
the various moments and compare the significance of running-coupling
effects to other perturbative contributions. We also quantify the effect
of the non-inclusive contribution at the next-to-leading order. 
Finally, we extract the leading power-corrections implied for the various
moments of $1\,-$ thrust by the ambiguity of the perturbative 
SDG result, and identify the regions of phase-space from which they emerge.
The conclusions are given in section 4.

\section{The thrust distribution in the single dressed gluon model}

The differential cross section with
respect to the thrust $d\sigma/dt$ can be calculated in
perturbation theory in the single dressed gluon (SDG) approximation as follows 
\beq
\left.\frac{d\sigma}{dt}(t)\right\vert_{\SDG}
=C_F\int_0^{1}\frac{d\epsilon}{\epsilon}\,\bar{a}_{\eff}
(\epsilon Q^2) \dot{\cal F}(\epsilon,t)=
C_F\int_0^{1}\frac{d\epsilon}{\epsilon}\,\bar{\rho}
(\epsilon Q^2) \left[{\cal F}(\epsilon,t)-{\cal F}(0,t)\right]
\label{res}
\eeq
where the integration over $\epsilon\equiv\mu^2/Q^2$ corresponds to
{\em inclusive} summation over final states into
which the emitted gluon fragments. In this calculation~\cite{DMW} 
a final state of invariant mass $\mu^2$ is characterized by a single universal
(i.e. observable independent) function $\bar{\rho}(\mu^2)$ which is 
identified as the time-like discontinuity of the
coupling $\a(k^2)\equiv \bar{\alpha}_s(k^2)/\pi$, where the bar stands for
a specific renormalization-scheme.
The two integrals in (\ref{res}) are related by integration by parts:
\beq
\dot{\cal F}(\epsilon,t)\equiv-\epsilon\frac{d}{d\epsilon}\,
{\cal F}(\epsilon,t),
\eeq
and the ``time-like coupling'' $\bar{a}_{\eff}(\mu^2)$ obeys 
\beq
\mu^2\,\frac{d\bar{a}_{\eff}(\mu^2)}{d\mu^2}=\rho(\mu^2). 
\eeq

The thrust characteristic function ${\cal F}(\epsilon,t)$ is obtained
from the following integral over phase-space,
\beq
{\cal F}(\epsilon,t)=\int_{\rm \tiny{phase\,\, space}} dx_1 dx_2 \, {\cal
M}(x_1,x_2,\epsilon) \, \delta\left(1-T(x_1,x_2,\epsilon)-t\right)
\label{F_def}
\eeq
where
$C_F\,a \,{\cal M}$ is the squared tree level matrix
element for the production of quark--anti-quark
pair and a gluon of virtuality
$\mu^2\equiv \epsilon Q^2$, and
\beq
{\cal M}(x_1,x_2,\epsilon)=\frac12\left[
\frac{(x_1+\epsilon)^2+(x_2+\epsilon)^2}{(1-x_1)(1-x_2)}
-\frac{\epsilon}{(1-x_1)^2}-\frac{\epsilon}{(1-x_2)^2}\right].
\label{M}
\eeq
The integration variables $x_{1,2}$ represent the energy fraction
of each of the quarks in the center-of-mass frame. The energy fraction
of the gluon is $x_3=2-x_1-x_2$.

For the calculation of the characteristic function we shall use the 
following~\cite{DMW,thrust} definition of the thrust 
\beq
T=\frac{\sum_i \left\vert \vec{p}_i \cdot \vec{n}_T\right\vert}
{\sum _i E_i }=\frac{\sum_i \left\vert \vec{p}_i \cdot \vec{n}_T\right\vert}
{Q}.
\label{T_def2}
\eeq
In case of three partons (a quark, an anti-quark  and a ``massive''
gluon) it yields~\cite{thrust},
\beq
1-T(x_1,x_2,\epsilon)={\rm min}\left\{ 1-x_1\,, \,1-x_2\,
,\,1-\sqrt{(2-x_1-x_2)^2-4\epsilon}
\right\}.
\label{thrust}
\eeq
Note that in the definition (\ref{T_def2}) the denominator 
is modified~\cite{DMW,thrust} with respect to the standard one (\ref{T_def}),
${\sum _i \vert \vec{p}_i \vert } \longrightarrow {\sum _i E_i }$, in a
way that does not change the observable for massless partons. 
The virtual (``massive'') gluon
is understood to fragment eventually into massless partons. This modification
ensures that the value of the thrust calculated with a ``massive'' gluon 
will be correct, provided that all the (massless) partons produced in
the process of the gluon fragmentation end up in the same hemisphere 
with respect to $\vec{n}_T$. 
The inclusive calculation performed here is justified only 
if the gluon fragmentation is predominantly collinear, in which case 
fragmentation into opposite hemispheres will be rare.
The discrepancy between the inclusive ``massive gluon'' calculation and the
full non-inclusive calculation was addressed before by several authors
\cite{Nason_Seymour,thrust,Dasgupta:1999mb,Beneke}. This issue will be further 
discussed in the next section in the context of the moments of
\hbox{$1-\,$ thrust}.

\begin{figure}[htb]
\begin{center}
\mbox{\kern-0.5cm
\epsfig{file=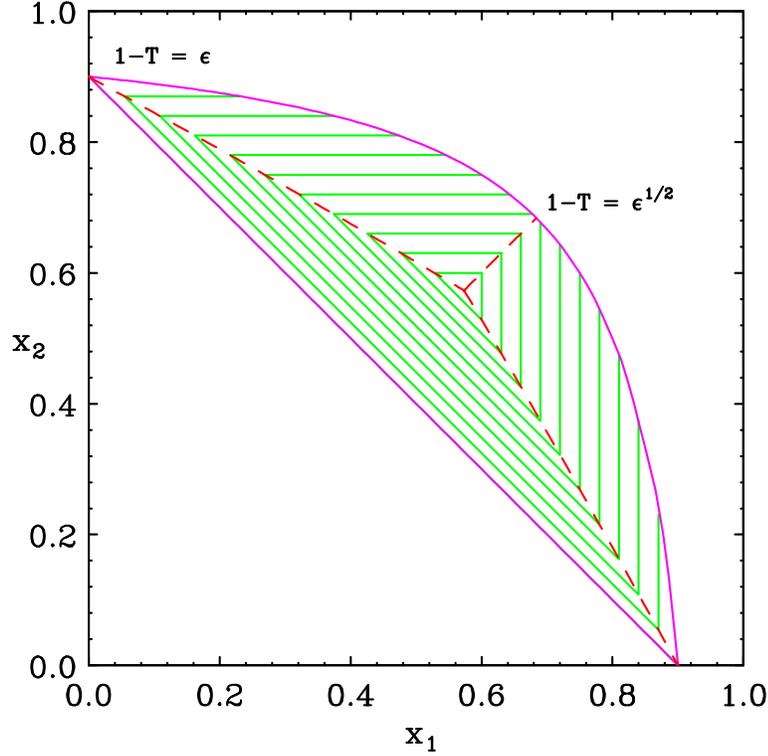,width=10.0truecm,angle=90}
}
\end{center}
\caption{Phase-space for the emission of a ``massive'' gluon with
 $\mu^2\,=\,\epsilon\,Q^2\,=\,0.1\,Q^2$
in the plane of the quark and anti-quark
energy fractions ($x_{1,2}$). The outer continuous lines represent phase-space
 limits, the dashed lines separate the phase-space according to 
the identification of the thrust axis $\vec{n}_T$ and internal 
continuous lines are constant thrust contours.
These contours are drawn with separation of $\Delta T=0.03$, 
stating at the lowest possible value of $1-T=\epsilon=0.1$ where two of 
the three partons are roughly collinear, up to the highest possible value of 
$1-T=  \left(2 \sqrt{1+3\epsilon}-1\right)/3\simeq 0.427$ where the momentum
splits equally between the three partons.
}
\label{phase_space}
\end{figure}

The last ingredient for the calculation of the characteristic function
is the phase-space\footnote{The reader is referred to~\cite{thrust} for
more details on the calculation of the phase-space boundaries.}. 
Fig.~\ref{phase_space} shows the three
parton phase-space in case of a gluon with a ``mass'' of 
$\mu^2\,=\,\epsilon\,Q^2\,=\,0.1\,Q^2$.
The external boundaries of phase-space in the figure correspond to 
the softest gluons (the upper curved line) and to the hardest ones 
(the lower linear line). 
The dashed lines represent the separation of phase-space according to 
which particle carries the largest momentum and thus determines the 
thrust axis (cf. eq.~(\ref{thrust})): in the upper left region $T=x_2$, 
in the upper right region $T=x_1$ and in the lower region 
$T=\sqrt{x_3^2-4\epsilon}$.

It is important to note that the two upper regions of phase-space in the
figure (where one of the primary quarks carries the largest 
momentum, $T=x_{1,2}$) have three corners with a definite
physical meaning for $\epsilon\ll 1$: 
\begin{description}
\item{(i) }
the collinear limit
corresponding to a two-jet configuration, with
$t=\epsilon$. This is the lowest possible value of $t$, given $\epsilon$.
\item{(ii) }
the three parton symmetric limit,
corresponding to a three-jet configuration, 
$t= \left(2 \sqrt{1+3\epsilon}-1\right)/3$. This is the highest possible
value of $t$, given $\epsilon$.
\item{(iii) }
the large-angle soft gluon limit, with $t=\sqrt{\epsilon}$.
Here gluons are emitted close to $90$ degrees with
respect to the quark--anti-quark direction.
\end{description}

It is clear from fig.~\ref{phase_space} that
in order to perform the integral in (\ref{F_def}) one has to treat 
separately values of $t$ smaller than $\sqrt{\epsilon}$ where the
soft gluon phase-space boundary (the curved line in fig.~\ref{phase_space}) 
is relevant, vs. $t$ larger than $\sqrt{\epsilon}$, where it is not.
As explained above, the value $t\simeq \sqrt{\epsilon}$ corresponds, for small
$\epsilon$, to the situation where the gluon is soft and emitted
in a large angle with respect to the quark--anti-quark direction. This
part of phase-space is the source of the $1/Q$ power-corrections for the 
average thrust, and as we shall see in the next section, 
also for the leading, though much
suppressed, infrared power-corrections for higher moments of $1\,-$ thrust, 
at the SDG level. 
Note that other limits of phase-space, as well as the squared matrix
element (\ref{M}) itself do not contain any $\sqrt{\epsilon}$ terms. 
In particular, in the collinear limit $t\simeq \epsilon$.

Evaluating (\ref{F_def}) we obtain the characteristic function for the
thrust distribution,
\begin{equation}
{\cal F}(\epsilon,t)= \left\{\begin{array}{lll}
{\cal F}_Q^l(\epsilon,t)+{\cal F}_G(\epsilon,t)&\,\,\,\,\,\,\,\,\,\,
& \epsilon<t<\sqrt{\epsilon}\\
{\cal F}_Q^h(\epsilon,t)+{\cal F}_G(\epsilon,t)&                  &\sqrt{\epsilon}<t<\frac23\sqrt{1+3\epsilon}-\frac13 
\end{array}\right.
\label{F_dist}
\end{equation} 
where the dominant contribution  ${\cal F}_Q(\epsilon,t)$ corresponds to
the phase-space regions where one of the primary quarks carries the
largest momentum ($T=x_{1,2}$) and ${\cal F}_G(\epsilon,t)$ corresponds
to the region where the gluon momentum is the largest (see 
fig.~\ref{phase_space}). The superscripts $l$ and $h$ on ${\cal
F}_Q(\epsilon,t)$ denote low and high $t$ values, respectively.
These functions are given by
\begin{eqnarray}
\label{F}
{\cal F}^h_{Q}(\epsilon,t) &=&  - \frac{1}{t}\left[(1 - 
t + \epsilon )^{2} + (1 + \epsilon )^{2}\right]\,\ln{ \frac {t}{q - t}}
+\left(3 - 2\,{ \frac {q}{t}}  + { 
\frac {1}{2}} \,{ \frac {1}{t}}  + { 
\frac {1}{2}} \,t - q\right) \nonumber \\
&+& \left(4 - 2\,{ \frac {q}{t}}  + 3\,
{ \frac {1}{t}}  - { \frac {q}{t^{2}}} 
 + { \frac {1}{q - t}} \right)\,\epsilon \nonumber \\
{{\cal F}^l_{Q}}(\epsilon,t) &=&  - \frac{1}{t}\left[(1
 - t + \epsilon )^{2} + (1 + \epsilon )^{2}\right]\,\ln
{ \frac {\epsilon }{t\,(q - t)}} 
+\left(1 - 2\,{ \frac {q}{t}}  + { \frac {1}{
2}} \,{ \frac {1}{t}}  - q\right) \nonumber \\
&+& \left(3\,{ \frac {1}{t}}  + { \frac {1}{q - t}}  - 
{ \frac {q}{t^{2}}}  + 2\,{ \frac {1}{t
^{2}}}  + 2 - 2\,{ \frac {q}{t}} \right)\,\epsilon   + 
\left(2\,+\, \frac {1}{2t}  \right)\,\frac{\epsilon ^{2}}{t^2}\\ \nonumber 
{\cal F}_{G}(\epsilon,t) &=&  
\frac{1 - t}{q^{2}} \left[ 
\left((1 + \epsilon )^{2} + (1 + \epsilon  - q)^{2}\right)
\ln\frac{q - t}{t} +(2t -q)\,\left(q + \frac {\epsilon}{t}  
 +\frac {\epsilon}{q-t} \right) \right] 
\end{eqnarray}
where $q \equiv \sqrt{(1 - t)^{2} + 4\,\epsilon}$.

Eq.~(\ref{res}) with ${\cal F}(\epsilon,t)$ given by (\ref{F_dist}) and
(\ref{F}) can now be used to calculate the SDG
contribution to the thrust distribution.
The leading order perturbative result \cite{Alt} 
can be recovered from (\ref{res})
by assuming a constant coupling $\bar{a}_{\eff} (\mu^2)\simeq {\rm
const.}$
This yields
\beq
\left.\frac{d\sigma}{dt}(t)\right\vert_{\rm LO}=C_F\,\bar{a}_{\eff} (\mu^2) 
\int_0^{1}\frac{d\epsilon}{\epsilon} \dot{\cal F}(\epsilon,t)=
C_F\,\bar{a}_{\eff} (\mu^2)  {\cal F}(0,t)
\label{LO}
\eeq
with
\beq
{\cal F}(0,t)=\frac{3t^2-3t+2}{t\,(1-t)}\,\ln\frac{1-2t}{t}-\frac32\frac{(1+t)(1-3t)}{t}
\label{F_LO}
\eeq
for $t\,\leq\,1/3$.
Eq.~(\ref{LO}) is the contribution to the thrust distribution of an on-shell
gluon. The improved calculation~(\ref{res}) 
takes into account the emission of ``massive'' 
gluons that later dissociate. For a given thrust value, the gluon virtuality
$\mu^2=\epsilon Q^2$ can change
in the following range:
\begin{equation}
\begin{array}{llr}
t<\frac13&\,\,\,\,\,\,\,\,\,\,& 0<\epsilon <t\\
t>\frac13&                    & \frac34t^2+\frac12t-\frac14<\epsilon <t
\end{array}
\label{t_values}
\end{equation} 
Whereas the leading order result (\ref{LO}) depends on an a priori
arbitrary choice of the renormalization scale $\mu^2$, eq.~(\ref{res})
resums running coupling effects to all orders, and is therefore
renormalization-group invariant\footnote{In fact, to have full
renormalization-group invariance we have to fix the scheme of $\bar{a}$.
This requires
further assumption concerning the diagrams that contribute to the gluon
fragmentation (``dressing the gluon''~\cite{thrust,conf}). Below we fix
it according to the Abelian-limit, in order to guarantee the correct
resummation of the terms leading in $\beta_0$.}. 
Note that performing the integral in 
eq.~(\ref{res}) can be in turn re-formulated as a specific choice 
of the renormalization point $\mu$ in (\ref{LO}) according to the 
BLM criterion~\cite{BLM}. For the thrust
distribution this representation seems rather cumbersome, since the
BLM scale will depend on $t$ in a complicated way. We shall use the BLM
formulation later on for the moments of \hbox{$1\,-$ thrust.} 

In spite of the resummation achieved by eq.~(\ref{res}), since this
calculation takes into account just a single gluon emission its
applicability is limited as follows: 
\begin{description}
\item{(i) } At ``high'' values $1\,-$ thrust, $t\,\gsim\,1/3$, only
non-planar configurations, e.g. more than three hadron jets, contribute.
The leading-order perturbative result (\ref{LO}) corresponding to three partons
vanishes identically for $t\,\geq\,1/3$ and the first non-vanishing 
contribution appears at the next order ${\cal O}(a^2)$. The resummation 
result~(\ref{res}) does not vanish above $t\,=\,1/3$. This fits the
intuitive expectation: the fragmentation of the ``massive'' gluon is 
not restricted to the three parton plane. On the other hand, it is clear 
that in this region emission of two hard gluons from the primary quarks is 
important.   
\item{(ii) } At low values of $1\,-$ thrust, $C_F\,a\,\ln^2(1/t)\,\gsim\,1$, 
the contribution of multiple emission of gluons which
are both soft and collinear becomes dominant and must be resummed to
all-orders in perturbation theory~\cite{CTTW}.
At even lower $1\,-$ thrust values,  $\beta_0\,a\,\ln(1/t)\,\simeq\,1$, 
this resummation breaks down since non-perturbative soft gluon emission
becomes important. 
\end{description}
In view of these facts, we can expect the SDG approximation to
describe the thrust distribution only in some restricted 
range of intermediate thrust values. Since soft and collinear gluon
resummation~\cite{CTTW} is complementary to dressing the gluon
(\ref{res}), the two resummation procedures can be combined to describe 
the thrust distribution in a wider range. 
The most promising approach to describe the differential
distribution of event-shape variables in the Sudakov region (here small $t$) 
is based on introducing a non-perturbative ``shape-function'' (SF)
\cite{Shape_function1,Shape_function2,Shape_function3,Shape_function4}. 
The physical distribution is then obtained by convoluting this
function with the perturbative result. 
Further analysis of the thrust distribution along these lines will 
appear in a separate publication. Here we proceed by analyzing 
moments of $1\,-$ thrust.
 
\section{Moments of $1\,-$ thrust}

Resummation of running-coupling effects in the SDG approximation was
already demonstrated to be important for the average thrust. In
particular, it was shown in~\cite{thrust} that the resummation modifies 
significantly the value of $\alpha_s$ extracted from experimental data. 
The application of the SDG approximation to the average thrust was
justified by asserting that low $t=1-T$ values are suppressed in 
the average, and therefore the effect of multiple soft and collinear gluons  
should not be significant. From this respect the calculation of higher 
moments of  $1\,-$ thrust, ${\large<} t^m {\large>}$, based on a SDG 
is even safer since they have a stronger suppression of the low $t$ region. 
On the other hand, the contribution to ${\large<} t^m {\large>}$ from
extremely high values of $1\,-$ thrust, $t\,\gsim\,1/3$, 
becomes more significant as $m$ increases. For high enough $m$ 
this contribution must be dominant, making the SDG calculation unreliable.   

Let us now assume that the contribution to ${\large<} t^m {\large>}$ is
dominated by intermediate values of $1\,-$ thrust where the SDG
approximation to the distribution (eq.~(\ref{res})) holds, 
and use it to calculate the $m$-th moment of $1\,-$ thrust according to 
\begin{eqnarray}
\label{res_mom}
{\large<} t^m {\large>}\sim
{\large<} t^m {\large>}_{\SDG}&=& 
\int\left.\frac{d\sigma}{dt}(t)\right\vert_{\SDG}t^m\,dt
=C_F\int\,t^m\,dt\int_0^{1}\frac{d\epsilon}{\epsilon}\,\bar{a}_{\eff}
(\epsilon Q^2) \dot{\cal F}(\epsilon,t) \nonumber \\
&=&C_F\int_0^{1}\frac{d\epsilon}{\epsilon}\,\bar{a}_{\eff}
(\epsilon Q^2) \dot{\cal F}_{{\large<}t^m{\large>}}(\epsilon),
\end{eqnarray}
where in the last step we changed the order of integration defining
\beq
{\cal F}_{{\large<}t^m{\large>}}\equiv \int {\cal F}(\epsilon,t)\,t^m\,dt.
\label{char_mom}
\eeq

In order to use the resummation formula (\ref{res_mom}) we have to
specify the time-like coupling $\bar{a}_{\eff}(\mu^2)$. 
As explained in~\cite{thrust,conf}, the renormalization-scheme defining
the coupling $\bar{a}$ should be uniquely determined once an
Abelian-like skeleton expansion is established in QCD. 
For the purpose of the current investigation we shall just use the 
one-loop approximation of the space-like coupling 
\beq
\bar{a}(k^2)=\frac{1}{\beta_0}
 \frac{1}{\ln\left(k^2/{\bar{\Lambda}}^2\right)}
\label{one-loop}
\eeq
with $\bar{\Lambda}$ set such that $\bar{a}$ coincides with the
Gell-Mann Low effective charge in the Abelian (large $\beta_0$)
limit. This is realized, for example, if $\bar{a}$ is related to the
$\overline{\rm MS}$ coupling by a scale-shift, i.e. 
\beq
\label{MSbar_rel}
\bar{a}(k^2)=\frac{a_{\MSbar}(k^2)}{1-\frac53\beta_0a_{\MSbar}(k^2)}.
\eeq
The corresponding time-like coupling is given by \cite{thrust}
\beq
\bar{a}_{\eff}(\mu^2)=\frac{1}{\beta_0}\,
\left[\frac{1}{2}-\frac{1}{\pi}\arctan\left(\frac{1}{\pi}
\log \frac{\mu^2}{\bar{\Lambda}^2}\right)\right].
\label{a_eff_oneloop}
\eeq
Using this coupling in (\ref{res_mom}), the terms which are leading in 
$\beta_0$ will be resummed correctly (ignoring the non-inclusive nature
of the observable \cite{Nason_Seymour,thrust,Dasgupta:1999mb}, which
will be discussed below), while other terms which are
sub-leading in $\beta_0$ will be neglected.

Before proceeding with the SDG analysis it is worthwhile to examine the
known next-to-leading order result \cite{Ellis}.
The next-to-leading order coefficient is calculated based
on numerical integration over the three and four parton phase-space 
\cite{EVENT,EVENT2}. In $\overline{\rm MS}$ it is given 
by~\cite{Biebel,Kluth_Biebel}
\begin{eqnarray}
\label{NLO_exact_coef}
{\large{<}}t{\large{>}}&=&0.7888\,{C_F}\, a_{\MSbar}(Q^2)+
 \left( - 1.1570\,{C_F}^2 + 4.4708\,
{C_F}\,{C_A} - 0.8445\,{C_F}\,{N_f}\right)\,a^{2}_{\MSbar}(Q^2)\nonumber\\
{\large{<}}t^2{\large{>}}&=&0.0713\,{C_F}\, a_{\MSbar}(Q^2)+
\left(0.3073\,{C_F}^2 + 0.3280\,
{C_F}\,{C_A} - 0.0583\,{C_F}\,{N_f}\right)\,a^{2}_{\MSbar}(Q^2)\nonumber\\
{\large{<}}t^3{\large{>}}&=&0.0112\,{C_F}\, a_{\MSbar}(Q^2)+
\left(0.06877\,{C_F}^2 + 0.04973\,
{C_F}\,{C_A} - 0.00808\,{C_F}\,{N_f}\right)\,a^{2}_{\MSbar}(Q^2)\nonumber\\
{\large{<}}t^4{\large{>}}&=&0.0022\,{C_F}\, a_{\MSbar}(Q^2)+
\left(0.01622\,{C_F}^2 + 0.00989\,
{C_F}\,{C_A} - 0.00145\,{C_F}\,{N_f}\right)\,a^{2}_{\MSbar}(Q^2) \nonumber\\
\end{eqnarray}

In order to correctly identify the terms that originate from the
\hbox{running-coupling~\cite{thrust,conf},} 
we use (\ref{MSbar_rel}) to translate
(\ref{NLO_exact_coef}) to the $\bar{a}$ scheme. Then, expressing 
the $N_f$ dependence of the next-to-leading order coefficients in terms of 
$\beta_0=(11/12)C_A-(1/6) N_f$ as done in the ``naive
non-Abelianization'' procedure~\cite{BBB} we obtain
\begin{eqnarray}
\label{NLO_exact_coef_bar}
{\large{<}}t{\large{>}}&=&C_F\left[0.7888\,\,\bar{a}(Q^2) + 
\left(3.7526\,\,\beta_{0} - 1.1567\,\,{C_F} - 
0.1740\,\,{C_A}\right)\,\bar{a}^{2}(Q^2)\right]\nonumber \\
{\large{<}}t^2{\large{>}}&=&C_F\left[0.0713\,\,\bar{a}(Q^2) + 
\left(0.2308\,\beta_{0} + 0.3073\,{C_F}+ 
0.00762\,\,{C_A}\right) \bar{a}^{2} (Q^2)\right]\\
{\large{<}}t^3{\large{>}}&=&C_F\left[0.0112\,\,\bar{a}(Q^2) + 
\left(0.02981\,\beta_{0} + 0.06877\,{C_F} 
+ 0.005271\,{C_A}\right)\,\bar{a}^{2}(Q^2)\right]\nonumber\\\nonumber
{\large{<}}t^4{\large{>}}&=&C_F\left[0.0022\,\,\bar{a}(Q^2) + 
\left(0.005045\,\beta_{0} + 0.01622\,{C_F} 
+ 0.00191\,{C_A}\right)\,\bar{a}^{2} (Q^2)\right]
\end{eqnarray}
Let us now look at the relative magnitudes of the different 
terms in the next-to-leading order coefficients in
(\ref{NLO_exact_coef_bar}).
Substituting the QCD values of $C_F$, $C_A$ and $\beta_0$ (for $N_f=5$)
we obtain the numerical values summarized in table \ref{NLO_coef_decomp_tab}.
\begin{table}[htb]
\[
\begin{array}{|c||c|c|c|c|}
\hline
&\,\,\,\,\,{\large<}t{\large>}\,\,\,\,\,&\,\,\,\,{\large<}t^2{\large>}\,\,\,\,
&\,\,\,\,{\large<}t^3{\large>}\,\,\,\,&\,\,\,\,{\large<}t^4{\large>}\,\,\,\\
\hline
\hline
\beta_{0}  &7.1925  &  0.4423 &   0.0571 & 0.00967 \\
{C_F}         &-1.542  &  0.4098 &   0.0917 & 0.02163 \\
{C_A}      &-0.522  &  0.0229 &   0.0158 & 0.00574 \\
\hline
\end{array}
\]
\caption{Decomposition of the next-to-leading order 
coefficient multiplying $C_F\bar{a}^2(Q^2)$ in
eq.~(\ref{NLO_exact_coef_bar}) for each of the moments 
of \hbox{$t\equiv 1-T$}. Here we substituted $C_A=3$, $C_F=4/3$ and~$N_f=5$.}
\label{NLO_coef_decomp_tab}
\end{table}
We find that the non-Abelian $C_FC_A$ component is always small.
The relative significance of the $C_F\beta_0$ term associated 
with the running-coupling decreases with $m$:
this term is absolutely dominant in the case of the
average thrust. It is still the largest in the case of 
${\large{<}}t^2{\large{>}}$, but it is no longer dominant. This trend
continues for higher moments ${\large <}t^m{\large >}$, $m=3$ and $4$.
On the other hand, the significance of the double emission term, 
proportional to ${C_F}^2$, increases with $m$. This fits the intuitive
expectation: higher moments of $1\,-$ thrust become sensitive to 
spherical configurations, and in particular, to multi-jet
configurations. This property, which makes the next-to-leading 
order perturbative series less reliable as $m$ increases, 
also implies that the significance of the resummation of
running-coupling effects by the SDG formula (\ref{res_mom}) compared to
other contributions decreases with $m$. Note, however, that with
increasing orders in perturbation theory, the terms associated with the
running-coupling are expected in general to increase fast and eventually 
dominate the coefficients. 

We now want to calculate the characteristic functions 
(\ref{char_mom}) for the first few moments.
Using eq.~(\ref{F_dist}) we have
\beq
{\cal F}_{{\large<}t^m{\large>}}(\epsilon) =
\int_{\epsilon}^{\frac23\sqrt{1+3\epsilon}-\frac13}
 {\cal F}_G(\epsilon,t)\,t^m\,dt
+\int_{\epsilon}^{\sqrt{\epsilon}}
 {\cal F}_Q^l(\epsilon,t)\,t^m\,dt
+\int_{\sqrt{\epsilon}}^{\frac23\sqrt{1+3\epsilon}-\frac13}
 {\cal F}_Q^h(\epsilon,t)\,t^m\,dt,
\label{char_mom_res}
\eeq
or, after rearranging the terms, 
\beq
{\cal F}_{{\large<}t^m{\large>}}(\epsilon) =
\int_{\epsilon}^{\frac23\sqrt{1+3\epsilon}-\frac13}
 \left[{\cal F}_G(\epsilon,t)+{\cal F}_Q^h(\epsilon,t)\right]\,t^m\,dt
-\int_{\epsilon}^{\sqrt{\epsilon}}
 \left[{\cal F}_Q^h(\epsilon,t)-{\cal F}_Q^l(\epsilon,t)\right]
\,t^m\,dt
\label{char_mom_res_1}
\eeq
where the difference
\[
{\cal F}^h_{Q}(\epsilon,t)-{\cal F}^l_{Q}(\epsilon,t)
 =  -  \frac{1}{t}\left[(1 - t + 
\epsilon )^{2} + (1 + \epsilon
)^{2}\right]\,\ln\frac{t^2}{\epsilon}
+ 2 + { \frac {t}{2}}  + \epsilon\left( 2- { \frac {2}{t^{2}}}\right)
 - \frac{\epsilon^{2}}{t^{2}} \left(  2  + \frac{1}{2t} \right)
\]
contains the essential information on the region $t\simeq
\sqrt{\epsilon}$. As we shall see below, non-analytic terms in the 
expansion of ${\cal F}_{{\large<}t^m{\large>}}(\epsilon)$ at small $\epsilon$ 
imply power-suppressed contributions to ${\large<}t^m{\large>}$. It is 
the $\sqrt{\epsilon}$ in the upper boundary in the second term of 
(\ref{char_mom_res_1}), corresponding to large-angle soft-gluon
emission, which is the source the leading non-analytic terms.

\begin{figure}[htb]
\begin{center}
\mbox{\kern-0.5cm
\epsfig{file=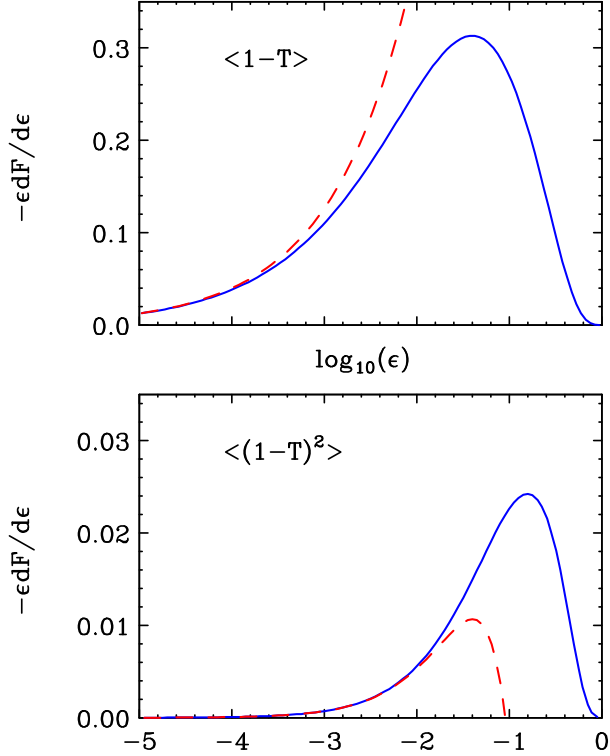,width=10.0truecm,angle=90}
}
\end{center}
\caption{The derivative of the characteristic functions 
$\dot{{\cal F}}(\epsilon)$
for ${\large<}1-T{\large>}$ (upper plot) and ${\large<}(1-T)^2{\large>}$ (lower plot) 
as a function of $\log_{10}(\epsilon)$,
where $\mu^2=\epsilon Q^2$ is the gluon virtuality. 
The dashed lines in the two plots represent the
${\cal O}(\epsilon^{1/2})$ and  ${\cal O}(\epsilon^{3/2})$
approximations to $\dot{{\cal F}}(\epsilon)$,
respectively.
}
\label{dF}
\end{figure}

\begin{figure}[htb]
\begin{center}
\mbox{\kern-0.5cm
\epsfig{file=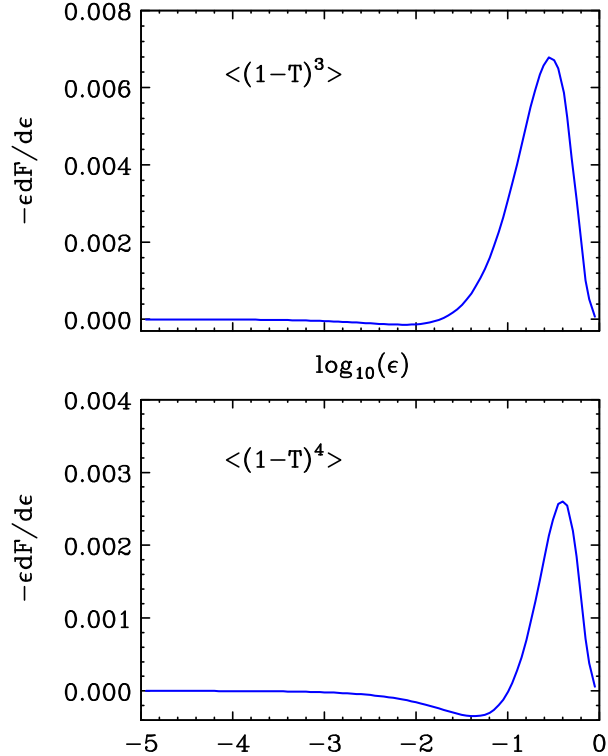,width=10.0truecm,angle=90}
}
\end{center}
\caption{The derivative of the characteristic functions 
$\dot{{\cal F}}(\epsilon)$
for ${\large<}(1-T)^3{\large>}$ (upper plot) and 
${\large<}(1-T)^4{\large>}$ (lower plot) 
as a function of $\log_{10}(\epsilon)$,
where $\mu^2=\epsilon Q^2$ is the gluon virtuality. 
}
\label{dF34}
\end{figure}

The resulting characteristic functions for $m=1$ and
$2$ are shown in fig.~\ref{dF} and these for $m=3$ and
$4$ -- in fig.~\ref{dF34}.
On this basis, the SDG perturbative sum in (\ref{res_mom}) can be calculated. 
As an alternative to evaluating the integral we can expand the 
coupling $\bar{a}_{\eff}(\epsilon Q^2)$ under the
integration sign, e.g. in terms for $\bar{a}\left(Q^2\right)$, obtaining, 
\beq
{\large<} t^m {\large>}_{\SDG}= C_F\left[d_0\,\bar{a}(Q^2)
+d_1\beta_0\,\bar{a}^2(Q^2)+ \left(d_2-\frac{\pi^2}{3}d_0\right)\beta_0^2
\,\bar{a}^3(Q^2)+\cdots\right],
\label{exp_res_mom}
\eeq
where the coefficients are expressed in terms of the log-moments of the
corresponding characteristic function, $d_0\equiv {\cal
F}_{{\large<}t^m{\large>}}(0)$ and
\beq
d_i\equiv
\int_0^{1}\dot{\cal F}_{{\large<}t^m{\large>}}
(\epsilon)\,\left(-\ln\epsilon\right)^i\,\frac{d\epsilon}{\epsilon}.
\label{d_i}
\eeq
The log-moments can be evaluated by a straightforward numerical
integration. The resulting values are summarized in table~\ref{coef_tab}. 
\begin{table}[htb]
\[
\begin{array}{|c||c|c|c|c|}
\hline
&\,\,\,\,\,{\large<}t{\large>}\,\,\,\,\,&\,\,\,\,{\large<}t^2{\large>}\,\,\,\,
&\,\,\,\,{\large<}t^3{\large>}\,\,\,\,&\,\,\,\,{\large<}t^4{\large>}\,\,\,\\
\hline
\hline
d_0&   0.7888\,   &   0.0713   &   0.0112   &   0.0022 \\
d_1&   3.588\,    &   0.1875   &   0.0160   &   0.000139\\
d_2&   21.06\,    &   0.6296   &   0.0218   &   -0.00903\\ 
d_3&   154.03\,  &   2.5922   &  -0.0020   &   -0.0555\\
d_4&   1368.38\,  &   12.721   &  -0.3254   &   -0.3108\\
d_5&   14464.59\, &   72.835   &  -3.1238   &   -1.8546\\
\hline
\end{array}
\]
\caption{First few log-moment of the characteristic functions 
for various moments of \hbox{$t\equiv 1-T$}.}
\label{coef_tab}
\end{table}
In the cases of the average thrust and ${\large<} t^2 {\large>}$ one
identifies already at the first few orders the characteristics of infrared
renormalons, namely fast increase (that asymptotically becomes
factorial) and a constant sign pattern. This
behavior is much delayed for the higher moments.

The approximation of ${\large<} t^m {\large>}$ by 
 ${\large<} t^m {\large>}_{\SDG}$ in~(\ref{res_mom}), 
or by (\ref{exp_res_mom}), can be improved in a straightforward manner by
matching it with the next-to-leading order 
result~(\ref{NLO_exact_coef_bar}). Such a procedure was used in~\cite{thrust}
for the phenomenological analysis of the average thrust.
The additional terms in~(\ref{NLO_exact_coef_bar}),
correspond primarily to double gluon emission (a ${C_F}^2$ term). 
They also include a non-Abelian $C_AC_F$ component, 
which depends~\cite{thrust,conf} on the identification of the 
coupling $\bar{a}$ beyond the Abelian limit, and finally, also some residual 
$\beta_0$ dependent piece, which is related~\cite{thrust} to the 
non-inclusive nature of the thrust.

Had the moments of the thrust been completely inclusive with respect 
to the fragmentation products of the gluon, the
terms proportional to $\beta_0$ should have been fully contained in the
 SDG result~(\ref{res_mom}) or~(\ref{exp_res_mom}). 
As explained in section~5.2 in
\cite{thrust}, in the inclusive case the log-moments 
$d_i$, for any $i$, are equal to the terms leading in $\beta_0$ 
in the exact calculation. 
In the present case the next-to-leading order coefficient $d_1$ 
calculated as a log-moment of the inclusive characteristic function 
differs from the actual $\beta_0$ dependent term in~(\ref{NLO_exact_coef_bar}).
The ``massive gluon'' inclusive treatment is justified
only if the discrepancy between them is small. In the case of 
${\large<} t {\large>}$ this discrepancy was found \cite{thrust} to be
 tiny. Comparing table~\ref{coef_tab} with
 eq.~(\ref{NLO_exact_coef_bar}) we find that it is only $4.4\%$! Evidently 
it increases for the higher moments of $1\,-$ thrust. 
For ${\large<} t^2 {\large>}$ 
the inclusive approximation still seems reasonable: the discrepancy is
 $18.7\%$. It is much worse for 
${\large<} t^3 {\large>}$, where the discrepancy is $46\%$ and it 
completely breaks down for ${\large<} t^4 {\large>}$.

The physical reason why the inclusive calculation for
 the high moments fails is that   
${\large<} t^m {\large>}$ becomes sensitive to large $t$ values.
The high $t$ region is correlated with high gluon virtuality (see 
e.g.~(\ref{t_values})). High virtuality allows
the gluon fragmentation products to spread in large angles compared to
the original gluon momentum. Consequently,
fragmentation into opposite hemispheres becomes more probable.
In such a case, a full non-inclusive calculation would yield a
higher thrust ($T$) value compared to the inclusive calculation that
takes into account the gluon momentum itself (this follows from 
(\ref{T_def2}) using the triangle inequality). 
As a result, the inclusive calculation under-estimates the 
thrust distribution $d\sigma/dt$ at large $t$ on the expense of
over-estimating it for somewhat lower $t$ values.
This is why for all the moments the full $\beta_0$ dependent term 
in~(\ref{NLO_exact_coef_bar}) is larger than the inclusive result~$d_1$.
The success of the inclusive approximation for the average thrust 
and for ${\large<} t^2 {\large>}$ is related to the fact that most of the 
gluon fragmentation is roughly collinear.

For the first two moments of $1\,-$ thrust, the coefficients 
in (\ref{exp_res_mom}) increase 
fast due to infrared renormalons already at the first few orders. 
It is clear that $\bar{a}(Q^2)$ is not
a good expansion parameter. A simple way to approximate the 
perturbative sum (ignoring the renormalon ambiguity) 
is the BLM method. In this case we approximate (\ref{res_mom}) by 
\beq
\label{res_mom_BLM}
{\large<} t^m {\large>}_{\SDG}\simeq
C_F\,\bar{a}_{\eff} (\mu^2_{\BLM})\,
\int_0^{1}\frac{d\epsilon}{\epsilon}\, 
\dot{\cal F}_{{\large<}t^m{\large>}}(\epsilon)=
C_F\,\bar{a}_{\eff} (\mu^2_{\BLM})\, {\cal F}_{{\large<}t^m{\large>}}(0)
\eeq
where, at leading order, the BLM scale is the center of the
characteristic function, i.e. the average virtuality of the gluon
contributing to ${\large<} t^m {\large>}_{\SDG}$, 
\beq
\mu_{\BLM}^2 = Q^2\exp\left(-
\int_0^{1}\dot{\cal F}_{{\large<}t^m{\large>}}(\epsilon)
\,\left(-\ln\epsilon\right)\,\frac{d\epsilon}{\epsilon}
\left/\int_0^{1}\dot{\cal F}_{{\large<}t^m{\large>}}
(\epsilon)\frac{d\epsilon}{\epsilon}\right.
\right)
\equiv Q^2 \exp\left(-d_1/d_0\right).
\label{mu_NLM}
\eeq
In principle, higher order log-moments (\ref{d_i}) can be used to
improve the approximation in (\ref{res_mom_BLM}) by further modifying
the scale (see, for example, eq.~(33) in \cite{conf}). For the
qualitative discussion here, the leading-order BLM approximation will suffice.
Using (\ref{MSbar_rel}) one can translate the BLM result to the 
$\overline{\rm MS}$ scheme,  
\beq
{\large<} t^m {\large>}_{\SDG}\simeq C_F\,a_{\MSbar} 
(\mu^2_{\BLM,\MSbar})\, {\cal F}_{{\large<}t^m{\large>}}(0)
\eeq
with $\mu^2_{\BLM,\MSbar}=\mu^2_{\BLM}\,e^{-5/3}$.

The first two log-moments in table~\ref{coef_tab} allow to calculate 
the leading-order BLM scales for the various moments of $1\,-$ thrust 
according to (\ref{mu_NLM}). The results are shown in table~\ref{BLM_tab}.
\begin{table}[htb]
\[
\begin{array}{|c||c|c|c|c|}
\hline
&\,\,\,\,\,{\large<}t{\large>}\,\,\,\,\,&\,\,\,\,{\large<}t^2{\large>}\,\,\,\,
&\,\,\,\,{\large<}t^3{\large>}\,\,\,\,&\,\,\,\,{\large<}t^4{\large>}\,\,\,\\
\hline
\hline
\mu_{\BLM}/Q&0.1028  & 0.2685  & 0.4889 & 0.9689\\
\mu_{\BLM,\MSbar}/Q&0.0447 &  0.1167 & 0.2125 & 0.4211 \\
\hline
\end{array}
\]
\caption{BLM scales in the ``skeleton scheme'' and in $\overline{\rm MS}$ 
for various moments of $t\equiv 1-T$.}
\label{BLM_tab}
\end{table}
The numbers in the first row of the table have the interpretation of
the average gluon virtuality contributing to ${\large<}t^m{\large>}$. We see
that 
\begin{description}
\item{(i) } the typical virtuality,
which is also the correct argument for the coupling in
(\ref{res_mom_BLM}), is much lower than $Q$ for the first moments of
$1\,-$ thrust, especially for the average thrust and for ${\large<}t^2{\large>}$. 
\item{(ii) } with increasing $m$ highly virtual gluons become dominant (at 
large $m$, \hbox{$\mu_{\BLM}\longrightarrow Q$}). 
\end{description}
These features can be read directly from the characteristic function
curves. In fig.~\ref{dF} we see that
 ${\cal F}_{{\large<}t{\large>}}(\epsilon)$ 
is wide and centered at very low virtualities while 
${\cal F}_{{\large<}t^2{\large>}}(\epsilon)$
is narrower and centered at higher virtualities. Fig.~\ref{dF34} shows
that this trend persists for the higher moments. Note that 
${\cal F}_{{\large<}t^m{\large>}}(\epsilon)$ for $m\geq 3$, 
as opposed to $m=1$ and $2$, is not positive definite. Thus for $m\geq 3$ the
interpretation of $\mu_{\BLM}$ as the average gluon virtuality contributing
to ${\large<}t^m{\large>}$ is no longer accurate. Note that for $m=4$ the
cancelation between positive and negative contributions to $d_1$ is
already large, making the approximation of the integral (\ref{res_mom})
by the coupling at the BLM scale absolutely unreliable. 

Let us now turn to discuss the power-correction implied by the SDG
perturbative sum. The starting point is the observation that 
eq.~(\ref{res_mom}), understood as an
all-order perturbative sum, is ill-defined due to infrared renormalons.
The standard way to resum asymptotic series of this type is the Borel
method. To obtain a Borel representation one expresses the coupling 
in (\ref{res_mom}) as
\beq
\bar{a}_{\eff}^{\PT}(\mu^2)\,=\,\frac{1}{\beta_0}\,\int_0^\infty \,dz\,
\bar{a}(z)\,\frac{\sin\,\pi z}{\pi z}\,
e^{-z\ln{\mu^2 \over{\bar{\Lambda}}^2}},
\label{aeff-borel}
\eeq
where the $\sin$ factor arises from the analytic continuation to the
time-like region.
For the one-loop coupling (\ref{one-loop}), $\bar{a}(z)=1$.
Substituting this in (\ref{res_mom}) and changing the order of
integration one arrives at the following Borel-sum, 
\beq
{\large<} t^m {\large>}_{\SDG}=\frac{C_F}{\beta_0}\int_0^{\infty}dz\,B_{{\large<}
t^m {\large>}}(z)\,e^{-z\ln\frac{Q^2}{{\bar{\Lambda}}^2}} 
\label{Borel}
\eeq
with 
\beq
B_{{\large<}t^m {\large>}}(z)=\frac{\sin\,\pi z}{\pi z}
\,\int_0^1\,\epsilon^{-z} \,\dot{\cal
F}_{{\large<}t^m{\large>}}(\epsilon)\,\frac{d\epsilon}{\epsilon}.
\eeq
The integral over $z$ in (\ref{Borel}) is ill-defined yielding an
ambiguous result (see \cite{thrust,Grunberg-pow} for more details). 
The physical reason for this ambiguity is that our
``perturbative'' calculation (\ref{res_mom}) actually depends on the coupling
at all scales, including the infrared. This ambiguity can thus be
resolved only at the non-perturbative level. Nevertheless, in practice
it is possible \cite{thrust} to cure the ambiguity by {\em defining} the 
perturbative sum through some regularization procedure, like the
principal-value or a momentum cutoff. At the next step, one adds to 
the regularized perturbative sum explicit power-suppressed terms that 
have the same dependence on $Q^2$ as the leading ambiguities and 
a normalization which is controlled by free parameters to be fixed by a fit.  
An attractive possibility \cite{DMW} is to write this 
parameterization in terms of the small $k^2$ moments of coupling 
$\a(k^2)$, assumed to be regular in the infrared at the non-perturbative 
level. Thanks to the assumed universality of $\a(k^2)$,
this parameterization immediately implies universality of the power
suppressed terms for different observables, up to calculable factors
which depend on the characteristic function. 

Notice that since the time-like coupling (\ref{a_eff_oneloop}) is finite 
for any $\mu^2$ and has an infrared
fixed-point: $\bar{a}_{\eff}(0)=1/\beta_0$, the integral
(\ref{res_mom}) with (\ref{a_eff_oneloop}) is well-defined. 
This is the so-called \cite{APT} ``Analytic Perturbation Theory'' (APT) 
result. It differs by ambiguous power terms\footnote{See 
\cite{thrust,Grunberg-pow} and refs. therein.} from the 
corresponding Borel-sum (\ref{Borel}). 
The APT result, just like the principal-value Borel-sum, is a specific 
regularization of the perturbative sum. 
Its advantage is that it is straightforward to calculate.

In general, if we ignore sub-leading perturbative terms that are not
included in the SDG approximation (\ref{res_mom}), the {\em physical} 
result should be given by the regularized perturbative 
sum plus power-corrections. The apparent ambiguity in the choice of the
regularization procedure is eliminated by the power terms. In absence
of relevant non-perturbative calculation, the only way to appreciate the
significance of the power terms is by comparing different regularization
procedures.  

One has to distinguish between two types of
power-suppressed ambiguities of the perturbative sum:
\begin{description}
\item{(i) } differences between various regularizations, e.g. between
the APT integral and the principal-value Borel-sum.
\item{(ii) } the specific differences which are associated with infrared
scales, i.e. space-like momentum scales at which the coupling is not 
under control in perturbation theory. 
\end{description}
This second type of ambiguity can be studied 
by introducing a momentum cutoff in the space-like 
region~\cite{Grunberg-pow,thrust}. It was shown to be related 
to {\em non-analytic} terms \cite{BBZ,BBB,DMW,Grunberg-pow,thrust} in the
small gluon virtuality (or $\epsilon$) expansion of the characteristic
function. In case of observables\footnote{This does not apply 
to moments of $1\,-$ thrust which cannot be expressed in terms of 
local operators.} 
that admit an operator product expansion (OPE), the parametric
dependence of these ambiguous terms on $Q$ should fit the dimensions of the 
higher-twist operators.   
On the other hand, the first type of ambiguity may appear due to 
both analytic and non-analytic terms in the small $\epsilon$ 
expansion of the characteristic function~\cite{Grunberg-pow}. 
This implies that certain
regularizations differ from others, as well as from the physical
non-perturbative result, by non-infrared non-OPE power terms.
The simplest possibility is that non-perturbative effects are restricted
to large distances. Then a momentum cutoff in the space-like region 
as introduced in~\cite{Grunberg-pow,thrust}, or
equivalently, the principal-value Borel-sum, define a class of favorable 
regularizations which differ from each other, as well as from the
physical result, just by infrared power-corrections. 
We stress that the APT integral does not belong to this class of
regularizations, and is suggestive of a different scenario for the
regularization of perturbation theory, advocated in~\cite{APT}. 

In order to study the ambiguities of perturbation theory 
that emerge from the SDG calculation 
in the case of moments of $1\,-$ thrust (\ref{res_mom}), we calculated 
the small~$\epsilon$ asymptotic expansions of the characteristic functions 
${\cal F}_{{\large<}t^m{\large>}}(\epsilon)$ given by 
eq.~(\ref{char_mom_res_1}). 
These expansions for $m=1$ through $4$ are summarized in the appendix in
eq.~(\ref{F_t}) through (\ref{F_t4}) and in table~\ref{asym_tab}. 
According to the explanation above, non-analytic terms in the expansion
of the characteristic functions are of special interest, since they
imply ambiguity of perturbation theory which is associated with large
distances.  There are two physically distinct sources of non-analytic terms 
contributing to the expansions (\ref{F_t}) through (\ref{F_t4}). 
The first is the contribution of collinear soft gluon emission. 
It originates in the lower integration limit ($t=\epsilon$) of the two terms
in~(\ref{char_mom_res_1}).
The second is large-angle soft-gluon emission, which originates in the 
upper integration limit ($t=\sqrt{\epsilon}$) of the second term 
in~(\ref{char_mom_res_1}).
To distinguish between these two regions of phase-space, we evaluated 
separately the second 
integral in (\ref{char_mom_res_1}) at its upper limit. 
The results, which represent the contribution of large-angle soft-gluon 
emission to ${\cal F}_{{\large<}t^m{\large>}}(\epsilon)$, are summarized 
in eq.~(\ref{F_t_soft}).
The expansions in the appendix can be used, based on the general 
formulae obtained in ref.~\cite{thrust}, to 
analyze the power-corrections implied by (\ref{res_mom}). 

Note first that all the leading non-analytic terms in the small
$\epsilon$ expansion of the characteristic functions are of the
half-integer type. These terms, as well as all the other half-integer
terms, originate exclusively in the large-angle
soft-gluon region through the upper limit $t=\sqrt{\epsilon}$ in the second
integral in (\ref{char_mom_res_1}). This conclusion follows from the 
comparison of eq.~(\ref{F_t_soft}) with the full expansion in 
eq.~(\ref{F_t}) through (\ref{F_t4}).

Note also that in eq.~(\ref{F_t}) through (\ref{F_t4}) there are no
logarithmic terms at order $\epsilon \ln\epsilon$. In general,
presence of such terms in the characteristic function implies
~\cite{DMW,thrust} infrared power-corrections that scale as $1/Q^2$. 
Their absence at the level of the moments of $1\,-$ thrust,
${\large<}t^m{\large>}$, is not
expected a priori, since they are present at the level of the thrust 
distribution, see eq.~(\ref{F_dist}) with (\ref{F}). 
Comparing eq.~(\ref{F_t_soft}) to eq.~(\ref{F_t}) it becomes apparent
that for the average thrust the $\epsilon \ln\epsilon$ terms cancel between
collinear and large-angle soft terms. For higher moments of $1\,-$  thrust
such terms do not appear even in the separate contributions of these two
phase-space regions (see eq.~(\ref{F_t_soft})).
The first non-vanishing logarithmic terms for the first two moments, 
$m=1$ and $2$, appear at order $\epsilon^2 \ln\epsilon$, namely $1/Q^4$.
These terms originate in both collinear and large-angle soft-gluon regions.
The leading logarithmic terms for higher moments, $m=3$ and $4$, 
appear much later, and exclusively\footnote{In eq.~(\ref{F_t_soft}) 
for $m=3$ and $4$ there are no logarithmic terms at
all.} due to the collinear limit. 
Since in non of the moments, do logarithmic terms appear as the leading
non-analytic terms we can safely ignore them in the more
detailed analysis that follows.

Let us now summarize some formulae~\cite{thrust}
for the regularization dependence of (\ref{res_mom}), using the
simplest one-loop ansatz (\ref{one-loop}) for the coupling.  
Given a generic term of the form\footnote{As stressed above we ignore
logarithmic terms. Regularization of integrals
like (\ref{res_mom}) in presence of such terms was also 
analyzed in~\cite{thrust}.} 
$c_n \epsilon^n$ in the small $\epsilon$ expansion of 
${\cal F}_{{\large<}t^m{\large>}}(\epsilon)
-{\cal F}_{{\large<}t^m{\large>}}(0)$, 
the difference between the APT regularization and the Borel-sum is given 
by \cite{BBB,Grunberg-pow,thrust},
\beq
\delta{\large<}t^m{\large>}\equiv -c_n \,\frac{C_F}{\beta_0} 
\,\left(\frac{\bar{\Lambda}^2}{Q^2}\right)^n
\, \exp(\pm i\pi n).
\label{delta_APT}
\eeq
Note that for half-integer values of $n$, this formula yields an ambiguous
imaginary result. This is the ambiguity of the Borel-sum. The
principal-value regularization amounts to taking the real part of the
Borel sum, and thus it coincides with the APT integral, as far as these
terms are concerned. On the other hand for integer values of $n$
the difference between the APT integral and the Borel-sum is unambiguous
and real.  

To quantify the ambiguities associated with infrared scales we define a
space-like momentum cutoff $\mu_I$, with
$t_I\equiv \ln\left(\mu_I^2/{\bar{\Lambda}}^2\right)$.
The contribution to the principal-value Borel-sum from
momentum scales below $\mu_I$ is given by \cite{thrust},
\beq
{\large<}t^m{\large>}_{\IR}\equiv -c_n\, \frac{C_F}{\beta_0}\,
 \left(\frac{\mu_I^2}{Q^2}\right)^n
\, \frac{\sin\, \pi n}{\pi}\, \exp(- nt_I)\,{\rm Ei}(nt_I).
\label{delta_IR}
\eeq 
In accordance with the general statement above,
${\large<}t^m{\large>}_{\IR}$ is non-zero for non-analytic terms with 
half-integer $n$ while it vanishes for analytic terms where $n$ is integer.

Using (\ref{delta_APT}) and (\ref{delta_IR}) with the expansions
of the characteristic functions in eq.~(\ref{F_t}) through~(\ref{F_t4}) 
we find the leading ambiguities associated with the perturbative 
sum~(\ref{res_mom}) of each moment of $1\,-$ thrust.
The two rows in table~\ref{Q_tab} correspond to 
(\ref{delta_APT}) and~(\ref{delta_IR}), respectively. 
The table shows the parametric dependence of the leading ambiguity
on the center-of-mass energy $Q$. 
Table~\ref{Mz_tab} shows the corresponding numerical values
of the power terms for $Q={\rm M_Z}$, normalized by the (approximate) 
perturbative result (\ref{res_mom_BLM}) with the BLM scales of 
table~\ref{BLM_tab}, for 
$\mu_I=2\,{\rm GeV}$, assuming $\alpha_s({\rm M_Z})=0.115$ and $N_f=5$. 
\begin{table}[htb]
\[
\begin{array}{|c||c|c|c|c|}
\hline
&\,\,\,\,\,{\large<}t{\large>}\,\,\,\,\,&\,\,\,\,{\large<}t^2{\large>}\,\,\,\,
&\,\,\,\,{\large<}t^3{\large>}\,\,\,\,&\,\,\,\,{\large<}t^4{\large>}\,\,\,\\
\hline
\hline
\delta {\large<}t^m{\large>}&1/Q &1/Q^2 & 1/Q^2& 1/Q^2 \\
{\large<}t^m{\large>}_{\IR}&1/Q &1/Q^3 & 1/Q^3& 1/Q^5 \\
\hline
\end{array}
\]
\caption{Leading ambiguities in the perturbative calculation 
for various moments of $t\equiv 1-T$}
\label{Q_tab}
\end{table}
\begin{table}[htb]
\[
\begin{array}{|c||c|c|c|c|}
\hline
&\,\,\,\,\,{\large<}t{\large>}\,\,\,\,\,&\,\,\,\,{\large<}t^2{\large>}\,\,\,\,
&\,\,\,\,{\large<}t^3{\large>}\,\,\,\,&\,\,\,\,{\large<}t^4{\large>}\,\,\,\\
\hline
\hline
\delta {\large<}t^m{\large>} /{\large<}t^m{\large>} &
\pm 0.075\, i& -0.38\,10^{-3}&0.22\,10^{-3} &0.37\,10^{-3}\\
{\large<}t^m{\large>}_{\IR} /{\large<}t^m{\large>} &
 0.17&0.14\,10^{-3}&-0.24\,10^{-3}&-0.35\,10^{-6} \\
\hline
\end{array}
\]
\caption{Relative magnitudes of the ambiguities in the perturbative 
calculation for various moments of $t\equiv 1-T$ at $Q={\rm M_Z}$.}
\label{Mz_tab}
\end{table}

As mentioned above, in all the cases the leading contributions
from the infrared (\ref{delta_IR}) appear due to half-integer terms 
associated with large-angle soft-gluon emission.
These ambiguities imply existence of non-perturbative power terms
which scale as $1/Q$ for ${\large<}t{\large>}$, as $1/Q^3$ for 
${\large<}t^2{\large>}$ and ${\large<}t^3{\large>}$, and as $1/Q^5$ 
for ${\large<}t^4{\large>}$.

Considering the leading ambiguity of the perturbative
sum (\ref{delta_APT}), a definite difference exists between the case 
of the average thrust, $m=1$, and higher moments, $m\geq 2$.
In the case of ${\large<}t{\large>}$ the leading ambiguity of the Borel-sum
is due to the same $\sqrt{\epsilon}$ term which dominates the infrared
contribution, whereas for ${\large<}t^m{\large>}$ with $m\geq 2$ the 
leading ambiguity originates in an {\em analytic}
term ($c_1 \epsilon$) in ${\cal F}_{{\large<} t^m{\large>}}(\epsilon)$. 
It is therefore not associated with large distance scales.
Moreover, a careful examination of the source of the terms proportional 
to $\epsilon$, (compare e.g. (\ref{F_t_soft}) with the full 
expansion~(\ref{F_t}) through~(\ref{F_t4})) shows that they
are not associated with any definite part of phase-space. 

The numerical values in table~\ref{Mz_tab} clearly indicate that all the
power-corrections that appear in the SDG approximation are small for 
any $m\geq 2$. 
The relative error at ${\rm M_Z}$ is less than a pro-mil. The case
of the average thrust is unique having a significant contribution
from the infrared: with the parameters quoted above it is 17\% (!) 
at~${\rm M_Z}$. We stress that in contrast with the average thrust
case, the difference between the APT and
Borel-sum regularizations for $m\geq 2$ becomes parametrically larger,
as well as numerically larger at ${\rm M_Z}$, than the infrared contribution
to the Borel-sum. This might give an opportunity to
use experimental data in order to constrain $1/Q^2$ power terms such as
the ones making the APT integral different from the principal-value Borel-sum.

\section{Conclusions}

We started in section 2 by calculating the characteristic function of
the thrust distribution, ${\cal F}(\epsilon,t)$. The result, 
summarized by eq.~(\ref{F_dist}) and~(\ref{F}), forms the basis for
resummation of running-coupling effects in the SDG approximation 
in this case. On its own the SDG approximation is expected to describe 
the physical
distribution only in a limited range of thrust values. However, when combined
with other techniques \cite{CTTW,Shape_function4}, the range of
applicability of the calculated distribution can be extended, leading
to more meaningful comparison with experimental data than available today.

In section 3 we concentrated on the first few moments of $1\,-$ thrust, 
${\large<}t^m{\large>}$, \hbox{$m=1$} through $4$. The corresponding characteristic
functions ${\cal F}_{{\large<}t^m{\large>}}(\epsilon)$ were obtained by a 
straightforward integration of ${\cal F}(\epsilon,t)$ and then used 
to study the properties of the SDG approximation to ${\large<}t^m{\large>}$.
We saw that the characteristic mass scale of gluons contributing 
to ${\large<}t^m{\large>}$ increases fast with $m$ (see table~\ref{BLM_tab}). 
Whereas typical gluon virtuality contributing 
to the average thrust is about $10\%$ of the center-of-mass energy $Q$, 
this fraction becomes $27\%$ for ${\large<}t^2{\large>}$. Still, if one is
using the $\overline{\rm MS}$ scheme, the natural renormalization scale
$\mu_{\BLM,\MSbar}$ is quite far from the naive choice $\mu=Q$:
for ${\large<}t^2{\large>}$ we have $\mu_{\BLM,\MSbar}\simeq 0.12\,Q$.
This means that the effect of the resummation compared to the naive
perturbative treatment is still very significant in the case of 
${\large<}t^2{\large>}$.

For high moments $m\geq 3$, a significant source of uncertainty in the
available next-to-leading order
perturbative approximation (see eq.~(\ref{NLO_exact_coef_bar})) 
is related to multi-jet configurations that 
contribute at high values of $t$, $t\gsim 1/3$. Improving the
approximation requires a full next-to-next-to-leading order calculation. 

Our renormalon analysis was performed in the inclusive ``massive gluon''
approach. The advantage of this approach, apart from its simplicity, is
that it naturally generalizes to include higher orders \cite{thrust} in
the $\beta$ function of the ``skeleton scheme''. In addition, it allows
for non-perturbative effects to be parametrized \cite{DMW} in a 
transparent way using an infrared regular running-coupling. 
Using this parameterization the universality assumption 
can be tested. 
Of course, the inclusive resummation is justified only if the non-inclusive
effect is small.
An alternative approach to perform renormalon resummation is based on the
``naive non-Abelianization'' procedure. 
One can define non-inclusive large $N_f$ characteristic functions (see 
\cite{BBM,Dasgupta:1999mb}) for ${\large<}t^m{\large>}$, and then
restore the full $\beta_0$ of the non-Abelian theory.
Comparing the expansion of the inclusive resummation formula
(\ref{res_mom}) to the next-to-leading order perturbative
expansion, we obtained a quantitative estimate of the non-inclusive
effect. The discrepancy at the next-to-leading order is very small ($4.4\%$)
for ${\large<}t{\large>}$ and it increases for higher moments.
In the case of ${\large<}t^2{\large>}$ the inclusive approximation 
still seems reasonable (a discrepancy of $18.7\%$) but less so for 
${\large<}t^3{\large>}$. Due to the appreciable discrepancy 
it is worthwhile to compute also the non-inclusive resummation. 

We find that within the framework of the SDG calculation,
power-corrections to ${\large<}t^m{\large>}$, with $m\geq 2$, are
highly suppressed (see tables~\ref{Q_tab} and~\ref{Mz_tab}).
The main question that remains open concerning the phenomenology of
${\large<}t^2{\large>}$ is the significance of 
power-corrections from configurations of three hard partons 
plus a soft gluon. Such corrections can be 
\cite{DW_dist,Shape_function3,Shape_function4,Beneke} 
as large as $\alpha_s(Q^2)/Q$. We further
address this subject below.

In the case of the average thrust the leading ambiguity of the
perturbative sum scales as $1/Q$. This ambiguity originates 
in a particular part of phase-space where a soft gluon is emitted at 
a large angle. Being associated with large distance physics, it is
quite clear that this ambiguity can be resolved only at the
non-perturbative level, by including explicit non-perturbative terms 
that fall as $1/Q$.
We showed that for higher moments of $1\,-$ thrust the infrared contribution
to the SDG perturbative sum is dominated by the same large-angle soft-gluon 
region of phase-space. The corresponding non-analytic terms in the
characteristic functions  
lead to much suppressed power-corrections, which scale as $1/Q^3$ for 
${\large<}t^2{\large>}$ and ${\large<}t^3{\large>}$ and as $1/Q^5$ for
${\large<}t^4{\large>}$. 

For the various moments of $1\, -$ thrust, ${\large<}t^m{\large>}$, 
as for any other time-like
observable, there are further ambiguities in the summation of 
perturbation theory that are not related to large distance scales. Such
ambiguities become apparent when comparing different regularizations,
such as the principal-value Borel-sum and the APT integral.
In the case of the moments of $1\, -$ thrust, these ambiguities scale 
as $1/Q^2$ and thus dominate over the infrared ambiguity for $m\geq 2$.
We saw that this type of ambiguity
is not associated to any particular part of phase-space. This suggest 
that, contrary to the soft emission ambiguities, it does not signal 
any genuine non-perturbative effects. It may be then possible to 
eliminate the ambiguity just by choosing the correct class of 
regularizations of the perturbative sum. The simplest possibility is
that this class is defined by a space-like momentum
cutoff~\cite{Grunberg-pow,thrust}, or equivalently by the
principal-value Borel-sum. One should be aware that other 
scenarios~\cite{APT} are possible as well.

Finally, we would like to view our conclusions concerning 
the power-corrections for the moments of $1\,-$ thrust in the SDG model
in the context of previous analysis of the thrust 
distribution. For this purpose we briefly recall the results 
of refs.~\cite{DW_dist,Shape_function3,Shape_function4}. 
Assuming a two-jet configuration, it was shown that the main effect 
of non-perturbative soft gluon emission is 
a shift of the Sudakov-resummed perturbative spectrum to higher values
of $t$: 
\beq
\frac{d\sigma}{dt}(t)=\left.\frac{d\sigma}{dt}\right\vert_{\PT}(t-\Delta t),
\label{shift}
\eeq
where $\Delta t=\lambda/Q$. 
A priori, this formula applies only in the range $\Delta t\ll t \ll 1/3$. 
It was demonstrated~\cite{DW_dist} that the measured thrust distribution
can be fitted over a large range of $t$ and $Q$ values by introducing such a
shift.

It is clear that~(\ref{shift}) cannot describe the physical distribution
at extremely small values of $t$, $t\lsim \Delta t$, 
where multiple non-perturbative soft gluon
emission becomes essential. This difficulty can be 
resolved~\cite{Shape_function3,Shape_function4}, by introducing a
non-perturbative (observable dependent) shape-function (SF) to describe the
energy flow in the final state. As explained in~\cite{Shape_function4},
in this case the physical distribution is obtained by a convolution of
the Sudakov-resummed perturbative spectrum with the shape-function. The
resulting distribution at extremely small $t$ (to the left of the
distribution peak) depends on the form of the shape-function but at 
higher $t$ it approximately coincides with the shifted 
distribution~(\ref{shift}). 

Both the simple shift model~\cite{DW_dist} and the shape-function 
approach~\cite{Shape_function3,Shape_function4}, like the
Sudakov-resummed perturbative spectrum on which they are based, strongly
rely on the two-jet kinematics. These approaches strictly do not apply
to the large $t$ region where one gluon becomes hard. 
The success of the fits~\cite{DW_dist,Shape_function3,Shape_function4}
in a large range of $t$ and $Q$ values is encouraging and it might suggest that
also the first few moments of $1\,-$ thrust could be studied in this framework.
Yet, one should be aware of the fact that the distribution is rather
flat at large $t$ (where one gluon becomes hard) and thus the total
effect of the shift, or the convolution with the shape-function, 
is minor there. On the other hand the first moments crucially depend 
on this very same region of $t$.  Therefore it is
dangerous to analyze the non-perturbative corrections to the moments 
relying on the apparent success of the fits to the distribution.

Nevertheless, it is interesting to see the consequences of the assumption
that the simple shift model~\cite{DW_dist} or the shape-function 
approach~\cite{Shape_function3,Shape_function4} apply beyond the small
$t$ region. In this case one 
finds~\cite{Shape_function3,Shape_function4,Webber:1999zj},
\begin{eqnarray}
\label{shift_effect}
{\large<}t{\large>}_{\SF}&=&{\large<}t{\large>}_{\PT}+\lambda_1/Q \\ \nonumber 
{\large<}t^2{\large>}_{\SF}&=&{\large<}t^2{\large>}_{\PT}
+2\lambda_1{\large<}t{\large>}_{\PT}/Q+\lambda_2/Q^2,
\end{eqnarray}
where the scales $\lambda_i$ are the moments of the shape-function and
in the case of a shift~(\ref{shift}), they are simply $\lambda_i=\lambda^i$. 
This can be compared to the infrared power-corrections found in the SDG
calculation (table~\ref{Q_tab}), namely
\begin{eqnarray}
\label{SDG_effect}
{\large<}t{\large>}_{\SDG}&=&{\large<}t{\large>}_{\PT}+\lambda/Q
 +{\cal O}(1/Q^3)\\ \nonumber 
{\large<}t^2{\large>}_{\SDG}&=&{\large<}t^2{\large>}_{\PT}+{\cal O}({1/Q^3}),
\end{eqnarray} 
baring the fact that the perturbative part ${\large<}t^m{\large>}_{\PT}$
can be quite different in the two approaches: in (\ref{shift_effect}) 
Sudakov resummation is implicitly assumed while in (\ref{SDG_effect}) 
SDG renormalon resummation is assumed. 

For the average thrust the two approaches predict the same type of
leading power-correction. Let us examine the case of ${\large<}t^2{\large>}$:
\begin{description}
\item{(i) } the leading non-perturbative term in (\ref{shift_effect}),
$2\lambda_1{\large<}t{\large>}_{\PT}/Q$, is proportional to $\alpha_s(Q^2)$.
Therefore, this term is attributed to soft emission around a 
configuration of three hard partons. Clearly, this is beyond the scope the 
SDG calculation performed here. On the other hand, this term
crucially depends on the extension of the
distribution models of~\cite{DW_dist,Shape_function3,Shape_function4} 
beyond their range of validity (two-jet kinematics). 
Therefore further theoretical work is required.
Note that this prediction for the leading power-correction of
${\large<}t^2{\large>}$ is rather 
easy to verify, or exclude, using experimental data: due to the large
ratio (see eq.~(\ref{NLO_exact_coef})) 
between the normalization of ${\large<}t{\large>}_{\PT}$ and
${\large<}t^2{\large>}_{\PT}$ in (\ref{shift_effect}), the
numerical significance of this power-correction is quite large.
\item{(ii) } the second term in (\ref{shift_effect}), $\lambda_2/Q^2$, 
is not suppressed by any power of $\alpha_s$, and 
is therefore attributed to soft emission
around the two-jet configuration. On the other hand, the SDG 
calculation~(\ref{SDG_effect}) does not yield any infrared $1/Q^2$ 
correction to ${\large<}t^2{\large>}_{\PT}$.
There are two possibilities how these facts can be reconciled. The first
is that once a more complete\footnote{The shape-function approach
resums the power-corrections to the distribution 
which are the most singular in the small $t$ limit, i.e. corrections
that scale as $1/(Q^nt^n)$ for any $n$. Less singular corrections to the
distribution, e.g. $1/(Q^{n} t^{n-1})$ can be quite important 
for the moments.}  
description of the thrust distribution is
achieved there will not be any infrared $1/Q^2$ correction to 
${\large<}t^2{\large>}$. The second is   
that an infrared $1/Q^2$ correction will emerge entirely from {\em double} 
gluon emission. 
It is a possible, but yet an unusual, situation that double soft 
gluon emission becomes parametrically less suppressed than a single 
soft gluon emission around the same hard configuration. Therefore it
will be interesting to investigate power-corrections 
to~${\large<}t^2{\large>}$ from double gluon
emission within the renormalon approach.
\end{description}
We comment that techniques for the systematic 
analysis of power-corrections in the full four parton phase-space 
are not well established yet. The discussion above clearly shows that
such techniques are necessary for the analysis of event-shape variables.

\acknowledgments{I am grateful to Yuri Dokshitzer, Georges Grunberg and
Gregory Korchemsky for very~interesting and useful discussions and to
Otmar Biebel for his great help.}

\newpage
\appendix
\section{Asymptotic expansions for the characteristic functions}

The asymptotic expansions of ${\cal F}_{{\large<}t^m{\large>}}(\epsilon)$,
calculated according to eq.~(\ref{char_mom_res}) 
or~(\ref{char_mom_res_1}) are the given in
eq.~(\ref{F_t}) through (\ref{F_t4}) below. Table~\ref{asym_tab}
summarizes the numerical values of the coefficients. 
\begin{eqnarray}
\label{F_t}
{\cal F}_{{\large<}t{\large>}}(\epsilon) 
&=& - 2\,\rm{dilog}\,3 - { \frac {1}{6}} 
\,\pi ^{2} - { \frac {1}{36}}  - { 
\frac {3}{8}} \,\ln\,3 \\ \nonumber &-& 4\,\sqrt{\epsilon } + \left[2 + 6
\,\ln\,3\right]\,\epsilon  - { \frac {80}{9}} 
\,\epsilon ^{3/2} \\ \nonumber
 &+& \left[{ \frac {28}{3}} \,\ln\,2 + 
2\,\rm{dilog}\,3 - { \frac {3}{2}} \,\ln\,
\left({ \frac {1}{\epsilon }} \right) + { 
\frac {17}{12}}  + { \frac {1}{6}} \,\pi ^{2}\right]\,
\epsilon ^{2} - 4\,\epsilon ^{5/2} \\ \nonumber
 &+& \left[{ \frac {8}{3}} \,\ln\,\left(
{ \frac {1}{\epsilon }} \right) + { 
\frac {14}{15}} \,\ln\,2 - { \frac {4}{5}} \right]
\,\epsilon ^{3} + \left[ - { \frac {64}{105}} \,
\ln\,2 - { \frac {11}{2}} \,\ln\,\left(
{ \frac {1}{\epsilon }} \right) + { 
\frac {2719}{1260}} \right]\,\epsilon ^{4}+\cdots
\end{eqnarray}
\begin{eqnarray}
\label{F_t2}
{\cal F}_{{\large<}t^2{\large>}}(\epsilon) 
&=& - 2\,\rm{dilog}\,3 - { \frac {1}{6}} 
\,\pi ^{2} - { \frac {9}{8}} \,\ln\,3 + 
{ \frac {17}{216}}  \\ \nonumber
 &+& \left[12\,\rm{dilog}\,3 + \pi 
^{2} - { \frac {91}{18}}  + { \frac {
127}{12}} \,\ln\,3\right]\,\epsilon  + { 
\frac {16}{9}} \,\epsilon ^{3/2} \\ \nonumber
 &+& \left[{ \frac {11}{6}} \,\pi ^{2} + 
{ \frac {1}{4}} \,\ln\,\left({ \frac {
1}{\epsilon }} \right) + 22\,\rm{dilog}\,3 - 13\,\ln\,3
 + { \frac {104}{3}} \,\ln\,2 + 
{ \frac {4}{3}} \right]\,\epsilon ^{2} + 
{ \frac {16}{9}} \,\epsilon ^{5/2} \\ \nonumber
 &+& \left[8\,\rm{dilog}\,3 + { \frac {188
}{15}} \,\ln\,2 - { \frac {1}{3}} \,\rm{
ln}\,\left({ \frac {1}{\epsilon }} \right) - { 
\frac {103}{30}}  + { \frac {2}{3}} \,\pi ^{2}\right]\,
\epsilon ^{3} \\ \nonumber
 &+& \left[{ \frac {5}{12}} \,\ln\,\left(
{ \frac {1}{\epsilon }} \right) - { 
\frac {1003}{315}}  + { \frac {152}{105}} \,\rm{
ln}\,2\right]\,\epsilon ^{4}+\cdots
\end{eqnarray}
\begin{eqnarray}
\label{F_t3}
{\cal F}_{{\large<}t^3{\large>}}(\epsilon) 
&=&- 2\,\rm{dilog}\,3 - { \frac {1}{6}} 
\,\pi ^{2} + { \frac {28}{135}}  - { 
\frac {83}{64}} \,\ln\,3 \\ \nonumber
 &+& \left[{ \frac {8}{3}} 
\,\pi ^{2} + { \frac {373}{16}} \,\ln\,3 - 
{ \frac {53}{9}}  + 32\,\rm{dilog}\,3\right]\,
\epsilon  - { \frac {4}{9}} \,\epsilon ^{3/2
} \\ \nonumber
 &+& \left[{ \frac {227}{12}}  + 
{ \frac {23}{6}} \,\pi ^{2} + 76\,\ln\,2 + 
46\,\rm{dilog}\,3 - { \frac {155}{4}} \,\rm{
ln}\,3\right]\,\epsilon ^{2} - { \frac {128}{225}} \,
\epsilon ^{5/2} \\ \nonumber
 &+& \left[ - { \frac {1127}{180}}  + 
{ \frac {302}{15}} \,\ln\,2 + \pi ^{2} + 12
\,\rm{dilog}\,3\right]\,\epsilon ^{3} - { \frac {4
}{9}} \,\epsilon ^{7/2} \\ \nonumber
 &+& \left[ - { \frac {1}{4}} \,\ln\,\left(
{ \frac {1}{\epsilon }} \right) - { 
\frac {34999}{5040}}  + { \frac {1201}{105}} \,
\ln\,2\right]\,\epsilon ^{4}+\cdots
\end{eqnarray}
\begin{eqnarray}
\label{F_t4}
{\cal F}_{{\large<}t^4{\large>}}(\epsilon) 
&=&- 2\,\rm{dilog}\,3 - { \frac {1}{6}} 
\,\pi ^{2} + { \frac {1259}{4860}}  - 
{ \frac {649}{480}} \,\ln\,3 \\ \nonumber
 &+&  \left[ - 
{ \frac {85}{9}}  + { \frac {3373}{80}
} \,\ln\,3 + 5\,\pi ^{2} + 60\,\rm{dilog}\,3\right]\,
\epsilon  \\ \nonumber
 &+& \left[{ \frac {4879}{108}}  + 
{ \frac {400}{3}} \,\ln\,2 + { 
\frac {5}{2}} \,\pi ^{2} - { \frac {326}{3}} \,
\ln\,3 + 30\,\rm{dilog}\,3\right]\,\epsilon ^{2} + 
{ \frac {32}{75}} \,\epsilon ^{5/2} \\ \nonumber
 &+& \left[ - { \frac {841}{30}}  - 
{ \frac {824}{15}} \,\ln\,2 - 
{ \frac {16}{3}} \,\pi ^{2} + 24\,\ln\,3 - 
64\,\rm{dilog}\,3\right]\,\epsilon ^{3} + { \frac {
32}{75}} \,\epsilon ^{7/2} \\ \nonumber
 &+& \left[ - { \frac {5849}{315}}  + 
{ \frac {372}{35}} \,\ln\,2 - 8\,\ln\,
3 - { \frac {8}{3}} \,\pi ^{2} - 32\,\rm{dilog
}3\right]\,\epsilon ^{4}+\cdots
\end{eqnarray}
\begin{table}[htb]
\[
\begin{array}{|c|l|c|c|c|c|}
\hline
\,\,\, {\rm term}\,\,\,&\,\,\,\rm {IR}\,\,\,
&{\large<}t{\large>}\,\,\,\,\,&\,\,\,\,{\large<}t^2{\large>}\,\,\,\,
&\,\,\,\,{\large<}t^3{\large>}\,\,\,\,&\,\,\,\,{\large<}t^4{\large>}\,\,\,\\
\hline
\hline
\epsilon^0&&+0.7888 & +0.0713 & +0.01120 &+0.002197\\
\hline
\epsilon^{1/2}&1/Q& - 4&&&\\
\hline
\epsilon^1& & +8.5917 & -0.7999 & +0.06557 & +0.019036\\
\hline
\epsilon^{3/2}&1/Q^3& -8.8889 & +1.7778 & -0.4444 & \\
\hline
\epsilon^2\,\ln\left(\frac{1}{\epsilon}\right)&1/Q^4& -1.5 & +0.25 & &\\
\hline
\epsilon^2&&+6.6575 & -2.4337 & +0.76777 & -0.215365\\
\hline
\epsilon^{5/2}&1/Q^5& -4 & +1.7778 &  -0.56889 & +0.426667\\
\hline
\epsilon^3\,\ln\left(\frac{1}{\epsilon}\right)&1/Q^6 
&+2.6667 & -0.3333 & &\\
\hline
\epsilon^3&& -0.1531 & +0.3399 & +0.32290& -0.429646\\
\hline
\epsilon^{7/2}&1/Q^7& & & -0.4444 & +0.426667\\
\hline
\end{array}
\]
\caption{Numerical values of the coefficients in the asymptotic 
expansions of the characteristic functions for various moments of 
$t\equiv 1-T$. For example, ${\cal F}_{{\large<}t{\large>}}(\epsilon)=0.7888-
4\,\sqrt{\epsilon}+8.5917\,\epsilon+\cdots$  }
\label{asym_tab}
\end{table}

Substituting only the upper limit $\sqrt{\epsilon}$ in the second 
integral in eq.~(\ref{char_mom_res_1}) we can isolate the large-angle
soft-gluon contribution to ${\cal F}_{{\large<}t^m{\large>}}(\epsilon)$. The
results are the following:
\begin{eqnarray}
\label{F_t_soft}
\left.{\cal F}_{{\large<}t{\large>}}(\epsilon)\right\vert_{\soft} 
&=& - 4\,\sqrt{\epsilon }
 - \epsilon \,\mathrm{ln}\left(\frac{1}{\epsilon}\right) 
- { \frac {80}{9}} \,\epsilon ^{3/
2} - \epsilon ^{2}\,\mathrm{ln}\left(\frac{1}{\epsilon}\right) 
 - 4\,\epsilon ^{5/2}\nonumber  \\ \nonumber 
\left.{\cal F}_{{\large<}t^2{\large>}}(\epsilon)\right\vert_{\soft}  &=&
  - \epsilon  + 
{ \frac {16}{9}} \,\epsilon ^{3/2}
 - \left[{ \frac {1}{4}} \,\mathrm{ln}\left(\frac{1}{\epsilon} 
\right) + { \frac {9}{4}} \right]\,\epsilon 
^{2} + { \frac {16}{9}} \,\epsilon 
^{5/2} - \epsilon ^{3}\\ 
\left.{\cal F}_{{\large<}t^3{\large>}}(\epsilon)\right\vert_{\soft}  &=&
- { \frac {4}{9}} \,
\epsilon ^{3/2} + { \frac {3}{4}} 
\,\epsilon ^{2} - { \frac {128}{225}
} \,\epsilon ^{5/2} + { \frac {3}{4
}} \,\epsilon ^{3} - { \frac {4}{9}} 
\,\epsilon ^{7/2}\\
\left.{\cal F}_{{\large<}t^4{\large>}}(\epsilon)\right\vert_{\soft}  &=&
 - { \frac {1}{4}} \,
\epsilon ^{2} + { \frac {32}{75}} \,
\epsilon ^{5/2} - { \frac {7}{18}} 
\,\epsilon ^{3} + { \frac {32}{75}} 
\,\epsilon ^{7/2} - { \frac {1}{4}
} \,\epsilon ^{4} \nonumber
\end{eqnarray}
Note that (\ref{F_t_soft}) presents exact results rather than asymptotic
expansions at small $\epsilon$.

\end{document}